\documentclass[sigconf]{acmart}
\usepackage{enumitem}
\AtBeginDocument{%
  }

\copyrightyear{2024}
\acmYear{2024}
\setcopyright{rightsretained}
\acmConference[MMASIA Workshops '24]{ACM Multimedia Asia
Workshops}{December 3--6, 2024}{Auckland, New Zealand}
\acmBooktitle{ACM Multimedia Asia Workshops (MMASIA Workshops '24),
December 3--6, 2024, Auckland, New
Zealand}\acmDOI{10.1145/3700410.3702114}
\acmISBN{979-8-4007-1314-9/24/12}
\begin{document}

\title{Reference-free automatic speech severity evaluation using acoustic unit language modelling}

\author{Bence Mark Halpern}
\email{halpern.bence.e8@f.mail.nagoya-u.ac.jp}
\email{halpernbence@gmail.com}
\affiliation{%
  \institution{Nagoya University}
  \city{Nagoya}
  \country{Japan}
}

\author{Tomoki Toda}
\email{tomoki@icts.nagoya-u.ac.jp }
\affiliation{%
  \institution{Nagoya University}
  \city{Nagoya}
  \country{Japan}
}

\renewcommand{\shortauthors}{Halpern et al.}

\begin{abstract} 
Speech severity evaluation is becoming increasingly important as the economic burden of speech disorders grows. Current speech severity models often struggle with generalization, learning dataset-specific acoustic cues rather than meaningful correlates of speech severity. Furthermore, many models require reference speech or a transcript, limiting their applicability in ecologically valid scenarios, such as spontaneous speech evaluation. Previous research indicated that automatic speech naturalness evaluation scores correlate strongly with severity evaluation scores, leading us to explore a reference-free method, SpeechLMScore, which does not rely on pathological speech data. Additionally, we present the NKI-SpeechRT dataset, based on the NKI-CCRT dataset, to provide a more comprehensive foundation for speech severity evaluation. This study evaluates whether SpeechLMScore outperforms traditional acoustic feature-based approaches and assesses the performance gap between reference-free and reference-based models. Moreover, we examine the impact of noise on these models by utilizing subjective noise ratings in the NKI-SpeechRT dataset. The results demonstrate that SpeechLMScore is robust to noise and offers superior performance compared to traditional approaches. \end{abstract}




\received[accepted]{28 October 2024}

\maketitle

\section{Introduction}

Speech severity evaluation is the task of automatically assigning a score to the speech of a pathological speaker, representing the severity of their speech impairments. This task holds significant importance, as currently it is done by speech language pathologists, which is subjective and time-consuming. The time needed to do these recordings attracts substantial economic costs, for example, in the Netherlands alone, we estimate the projected annual increase in related healthcare costs of at least one million EUR if no action is taken \cite{hollandzorg2023}. Automating this process, and efficient triaging of patients is vital to reduce time and costs.

Recent advancements in automatic speech recognition (ASR) technology, particularly for typical speech, have propelled speech severity evaluation forward. Several approaches have shown good performance on speech intelligibility evaluation using word accuracy \cite{maier2009peaks}, and phonological features \cite{middag2009automated}. These approaches all rely on the phenomenon that ASR models make mistakes in the speech of pathological speakers.

One of the most pressing issues of these ASR-based approaches is that they often require a spoken or written reference, which restricts the evaluation of speech severity to read speech corpora. Read speech, however, is not a realistic representation usually of the speaker's real-life speech usage. This lack of realisticness is also called the lack of ecological validity in the literature.

Consequently, recent research has focused on ASR- and reference-free approaches to speech severity evaluation. Supervised learning reference-free approaches have been demonstrated to fail to learn meaningful features of the speech itself, instead relying on shortcuts embedded in the dataset, which compromises their effectiveness \cite{schu2023using, liu2024clever}. On the other hand, unsupervised approaches for speech severity evaluation almost exclusively consist of hand-crafted acoustic features such as jitter, shimmer and $F_{0}$ statistics \cite{tsanas2009accurate, tsanas2012novel, haderlein2017robust}. While these approaches have the advantage of providing ease of interpretability to the results, their success is mostly limited to idealised acoustic conditions and non-spontaneously elicited speech.

Recently, an interesting line of research inspired by text-to-speech synthesis techniques has repeatedly shown that human listeners have difficulty differentiating between the naturalness of the speech (i.e., how easy it is to tell apart from computer synthesised speech) and the severity of the speech (i.e., the level/extent of the speech impairment) \cite{huang2022towards, halpern2023improving, illa2021pathological}. It has been also shown that the scores of automatic naturalness evaluation methods highly correlate with the scores of automatic severity evaluation \cite{halpern2023automatic}. Given the tremendous effort invested in automatic naturalness evaluation \cite{huang2022voicemos, huang2024voicemos, cooper2023voicemos}, it follows that adoption of naturalness evaluation approaches could be useful for severity evaluation.

Inspired by this, we adopt a reference-free method called SpeechLMScore \cite{maiti2023speechlmscore}, a naturalness evaluation approach which does not rely on training with pathological speech databases. We demonstrate the approach has superior performance over existing acoustic feature approaches, and is robust to noise.

Additionally, we present the NKI-SpeechRT dataset, an extended version of the NKI-CCRT dataset used in previous research, providing a more comprehensive foundation for speech severity evaluation. As the dataset also contains subjective noisiness scores along with ratings of other speech features, this dataset also allows us to evaluate the robustness of the features to noise.

Our research questions are as follows:
\begin{enumerate}[label=\textbf{RQ\arabic*},noitemsep]
    \item Does SpeechLMScore perform better than the acoustic features approaches used previously in the literature?
    \item If it performs better, how large is the gap in performance between reference-free and reference-based models?
    \item Are SpeechLMScore and the compared acoustic models influenced by noise present in the recordings?
\end{enumerate}

\section{Datasets}

\subsection{NKI-OC-VC}

The NKI-OC-VC dataset \cite{halpern2023improving} includes Dutch pathological speech from 16 oral cancer (OC) speakers (10 male, 6 female) who had undergone a composite resection (COMANDO) surgery or comparable treatment for mostly advanced tongue tumours.

For six patients (four male, two female), data was collected from the participants at a maximum of three time points: before the surgery, within a month after the surgery, and approximately six months after surgery. The recordings took place during scheduled speech therapy sessions. Participants were asked to read the Dutch text ``Jorinde en Joringel'' \cite{son01_eurospeech} consisting of 92 sentences during the recording session. The total duration of all speech recordings, across all speakers, was approximately 2.5 hours. One recording session (speaker/time point) lasted five minutes on average. In some cases, patients felt the experiment was too burdensome, in that case, we prematurely stopped the experiment.

The speech was recorded with a Roland R-09HR field recorder at 44.1 kHz sampling frequency and 24-bit depth. This was later downsampled to 16 kHz and quantized to 16-bit.  The dataset includes speech severity labels provided by five speech language pathologists (SLPs) using a five-point Likert scale with 5 meaning healthy, and 1 meaning severe.  The interrater correlation between the intelligibility scores were very high, so the scores are more than reliable for further analysis ((\textit{ICC 2,k})=0.9671).

\subsection{NKI-SpeechRT}

We derive a dataset from the NKI-CCRT dataset for the task of speech severity evaluation. The dataset contains 55 speakers in total, with 45 male and 10 female speakers. Only 47 speakers are native Dutch speakers.  Participants were asked to read the Dutch text 'De vijvervrouw’ by Godfried Bomans. 
Recordings were made with a
Sennheiser MD421 Dynamic Microphone and portable 24-
bit digital wave recorder (Edirol Roland R-1). The speech samples were all downsampled to 16 kHz and quantized to 16-bit for later analysis. 

The dataset includes recordings from the speaker from a maximum of five stages of treatment, including before CCRT, 10 weeks post-CCRT, and 12 months post-CCRT. In total, 192 speaker-stage time points are included in the evaluation.

A speech evaluation experiment was carried out online after the recordings.
In the 70-minute online listening test, 14 Dutch recent SLP graduates without hearing difficulties rated the entire speech stimuli cut into three, approximately equal length segments. The audio was presented at 70 dB using Sennheiser HD418 headphones, and participants were able to see the text with the ability to replay the stimuli. Several dimensions such as the voice quality, intelligibility, and accentedness were rated on a 7-point Likert scale. In the current work, we have only used intelligibility. The interrater correlation between the intelligibility scores was very high, so the scores are more than reliable for further analysis ((\textit{ICC 2,k})=0.9174).

In practice, intelligibility achieved a high correlation with voice quality features, therefore, we do not think this has any impact on the evaluation. For a more detailed explanation of the experiment conditions, we refer the reader to Clapham et al's work \cite{clapham2012nki, clapham2014developing}.

In the case of clinically recorded data, it is unfortunately well known that recordings can highly vary due to various issues. To mitigate this, noise scores have also been collected for the dataset on a separate occasion. A non-SLP linguist was asked to provide noisiness scores for the dataset on a 3-point scale from 0 to 2. Zero meant no or barely audible disturbances, one meant audible disturbance, and two meant noisy disturbances including sometimes other voices or ringing of the telephone. For reference, the correlation between the noise and intelligibility annotations was -0.1435.

\section{Methods}

As there is a considerable amount of acoustic feature-based approaches existing in the literature, we were only capable of comparing to a small selection of acoustic features. We prioritised acoustic features which had publicly available implementations in Python. In the following sections, we will briefly explain the acoustic measures used, and justify their choice.

\subsection{Speaker-level experimental design}

In order to compare the different methods for speech severity evaluation, we take all the utterances of the speaker, and obtain an estimate of the utterance severity $\hat{x} \in \mathbf{R}$ by using the various approaches (acoustic features) introduced below.  In the case of short-time acoustic features, i.e. when the feature is a time-series, we calculate the mean to obtain a single scalar. Finally, we calculate the correlation of the mean of the utterance level features $\bar{\hat{x}}$, and the perceptual scores, and report it. Therefore, we are using a speaker-level severity evaluation in this work.

\subsection{Baseline approaches}

\textbf{Shimmer} refers to the variation in amplitude between consecutive voice cycles, commonly used to assess vocal instability.

\begin{equation*}
\hat{x}_{\text{shimmer}} = \frac{1}{N-1} \sum_{i=1}^{N-1} \left| \frac{A_{i+1} - A_i}{A_i} \right|
\end{equation*}

\textbf{Jitter} is often viewed as the pair of shimmer, which measures the irregularity in frequency between cycles, often indicating vocal pathologies. 

\begin{equation*}
\hat{x}_{\text{jitter}} = \frac{1}{N-1} \sum_{i=1}^{N-1} \left| \frac{T_{i+1} - T_i}{T_i} \right|
\end{equation*}

Shimmer and jitter have been historically often used for evaluating pathological speech, for example in Parkinson's speech \cite{tsanas2009accurate, tsanas2012novel, haderlein2017robust}.

\textbf{$\sigma{F_{0}}$} is the standard deviation of fundamental frequency, which has been extensively used for the blind estimation of severity in dysarthric speech \cite{paja2012automated, falk2012characterization}, as it has been shown that dysarthric speakers tend to demonstrate a smaller variation in $F_{0}$ \cite{bunton2000perceptuo}. 

\begin{equation*}
\hat{x}_{\sigma F_0} = \sqrt{\frac{1}{T} \sum_{t=1}^{T} (F_0(t) - \bar{F_0})^2}
\end{equation*}

\textbf{Voicing ratio} is the proportion of voiced sound frames to all the sound frames. The voicing ratio has also been used in many works, including for the evaluation of dysarthria \cite{paja2012automated} and for laryngectomy speech \cite{van2019acoustic}.

\textbf{Harmonics to noise ratio} (HNR) quantifies the degree of periodicity in the signal, helping differentiate healthy voices from pathological ones \cite{boersma1993accurate}. It has been found useful for a different version of the corpus when used in a supervised setting \cite{fang2017intelligibility}, and also in the speech of individuals with Parkinson's disease.
All of the above features have been estimated using \texttt{praat-parselmouth} library.

\textbf{WADA SNR} is one of the standard implementations of non-instrusive signal-to-noise measures \cite{kim2008robust}. We use a publicly available python implementation \footnote{https://gist.github.com/johnmeade/d8d2c67b87cda95cd253f55c21387e75}. Signal-to-noise-ratio is a correlate of speech problems, for example in oral and laryngeal cancer \cite{zhang2008acoustic, woisard2022construction}.

\textbf{CPP} (Cepstral Peak Prominence) evaluates voice quality by measuring the harmonic structure, particularly breathiness \cite{fraile2014cepstral}. CPP has been used in Parkinson's speech for example \cite{haderlein2011intelligibility}. The implementation provided in this repository is used \footnote{https://github.com/satvik-dixit/CPP}.

\subsection{Proposed approach: SpeechLMScore}

SpeechLMScore measures how likely a speech sample is to resemble natural speech by using a pretrained speech-unit language model, as described in \cite{maiti2023speechlmscore}. In our setup, we use a pretrained \texttt{HUBERT-BASE-LS960H} model to extract self-supervised speech representations. Given a speech utterance \( x_t \) at time \( t \), the model outputs hidden representations \( h_t \), where \( h_t = \text{HuBERT}(x_t) \). These representations are then quantized using k-means clustering, mapping each \( h_t \) to discrete acoustic tokens \( d_t \in \{1, \dots, K\} \), where \( K \) is the total number of clusters.

For the language modelling component, we use an LSTM trained on the LibriLight dataset \cite{kahn2020libri} to predict the next acoustic unit based on the sequence of previously observed units. The model assigns a probability \( p(d_t | d_{<t}) \), where \( d_{<t} \) denotes the sequence of prior acoustic units. We experimented with different layers of \texttt{HUBERT-BASE-LS960H} and found that layer 1 provided the most informative representations for speech severity.

Finally, the perpelexity is calculated, where lower perplexity values suggest that the model finds the sequence more natural. In our experiments, we use this perplexity value, \( \hat{x} \), as a direct correlate of the severity of speech impairment.

% More information is needed about layers, dataset

\subsection{Reference-based upper bound}

We use the phoneme error rate (PER) to provide a reasonable upper-bound for the reference-free experiments. We use a publicly available implementation of a CTC-based phoneme recogniser \footnote{\url{https://huggingface.co/Clementapa/wav2vec2-base-960h-phoneme-reco-dutch}}. The \texttt{facebook/wav2vec2-base-960h} base model was used for training on the Dutch partition of the Common Voice dataset \cite{ardilacommon}. We used \textit{phonemizer} to acquire phonetic transcriptions from the ground truth grapheme-level transcriptions provided in the dataset \cite{Bernard2021}. 

\section{Results and discussion}

\subsection{Research question 1: Performance of SpeechLMScore}

Table \ref{tab:correlation} shows that SpeechLMScore outperforms the traditional acoustic feature-based approaches across both datasets. In the NKI-SpeechRT dataset, SpeechLMScore achieved the highest correlation with listener ratings (r = 0.3834, p < 0.001), with only HNR being the second best (r = -0.2999, p < 0.001), and WADASNR being the third best (r = -0.2852, p < 0.001).

Similarly, in the NKI-OC-VC dataset, SpeechLMScore showed an even stronger correlation (r = 0.6895, p < 0.001), with WADA SNR being second best (r = -0.6350, p < 0.001), and jitter coming as third (r = 0.4528, p < 0.001)

The superior performance of SpeechLMScore is not surprising given the fact that it is a significantly more complex, and larger feature than the other acoustic features complicated. Another explanation for the superior performance of SpeechLMScore over the other acoustic measures is the fact the oral cancer speech patients in the corpora can be roughly said to have articulatory issues, while the acoustic features mainly concern changes in voice quality. Using articulatory measures, such as the ones \cite{tienkamp2023objective} would have been more sensible, however, these require precise segmentations for certain phonetic features which is difficult to acquire automatically for speakers with high severity.

It is also important to discuss the performance gap between the two datasets, i.e. both the SpeechLMScore and the Phoneme Error Rate have better performance on the NKI-OC-VC datasets. We think this difference can be owed to the fact that (1) the NKI-SpeechRT contains a broader range of voicing problems while the NKI-OC-VC contains mainly articulation (2) the differences in the rating schemes for the two tasks, i.e., 7-point scale was used instead of a 5-point scale, and more raters were used to obtain the mean scores for the NKI-SpeechRT.

\subsection{Research question 2: Performance gap between reference-free and reference-based}
The gap in performance between the reference-free SpeechLMScore and reference-based models, such as phoneme error rate (PER), varies across the datasets but remains significant. In the NKI-SpeechRT dataset, the correlation for SpeechLMScore (r = 0.3834, p < 0.001) is lower than the strong negative correlation achieved by the reference-based phoneme error rate (r = -0.8206, p < 0.001). However, in the NKI-OC-VC dataset, SpeechLMScore (r = 0.6895, p < 0.001) approaches the performance of the phoneme error rate (r = -0.9155, p < 0,001). While the reference-based model still performs better, the gap is narrower in NKI-OC-VC, demonstrating that SpeechLMScore is a lucrative alternative, especially considering that it does not require a reference transcript.

\begin{table}[ht]
\centering
\begin{tabular}{lcc}
\toprule
\textbf{Feature}      & \textbf{NKI-SpeechRT}           & \textbf{NKI-OC-VC}         \\ \midrule
\textbf{Shimmer}       & 0.1475 (0.0843)             & -0.1334 (0.4744)           \\ 
\textbf{$\sigma{F{0}}$}      & -0.1710 (*)            & 0.3208 (0.0785)            \\ 
\textbf{Jitter}        & 0.1257 (0.1417)             & 0.4528 (*)            \\ 
\textbf{WADA SNR}          & -0.2852 (***)            & -0.6350 (***)           \\ 
\textbf{Voicing\%} & 0.0273 (0.7506)             & -0.1768 (0.3413)           \\ 
\textbf{HNR}   & -0.2999 (***)            & 0.1355 (0.4675)            \\ 
\textbf{CPP}           & -0.1562 (0.0674)            & -0.2666 (0.1472)            \\ 
\textbf{SpeechLMScore}    & \textbf{0.3834} (***)           & \textbf{0.6895} (***)          \\ 
\midrule
\textbf{Phoneme Error Rate}           & -0.8206 (***)          & -0.9155 (***)         \\ 
\bottomrule
\end{tabular}
\caption{Pearson's correlation of listener scores with automatic scores of the respective acoustic feature or system across the NKI-SpeechRT and NKI-OC-VC datasets. P-values are written in parentheses. Smaller than 0.05 (*), 0.01 (**), 0.001 (***), otherwise full p-value.}
\label{tab:correlation}
\end{table}

\begin{table}[ht]
\centering
\begin{tabular}{lcc}
\toprule
\textbf{Feature}      & \textbf{NKI-SpeechRT (r)}    \\ 
\midrule
\textbf{Shimmer}               & 0.1620 (*)               \\ 
\textbf{$\sigma{F{0}}$}             & 0.0894 (0.2175)            \\ 
\textbf{Jitter}                & -0.0004 (0.9953)      \\ 
\textbf{WADA SNR}                  & -0.2461 (***)  \\ 
\textbf{Voicing\%}         & -0.1708 (*)      \\ 
\textbf{HNR}          & -0.0092 (0.8996) \\ 
\textbf{CPP}                   & 0.1596 (*)           \\ 
\textbf{SpeechLMScore}            & 0.0305 (0.6741)              \\ 
\midrule
\textbf{Phoneme Error Rate}      & 0.1459 (*)           \\ 
\bottomrule
\end{tabular}
\caption{Pearson's correlation of noise scores with automatic scores of the respective acoustic feature or system across in the NKI-SpeechRT. P-values are written in parentheses. Smaller than 0.05 (*), 0.01 (**), 0.001 (***), otherwise full p-value.}
\label{tab:noise_results}
\end{table}

\subsection{Research question 3: Noise influence}

The results in Table \ref{tab:noise_results} indicate varying degrees of correlation between acoustic features and the noise scores, with lower absolute correlations being more desirable as they suggest a reduced influence of noise. Jitter (r=-0.0004 , p = 0.9953), harmonics-to-noise ratio (r=-0.0092, p = 0.8996), and SpeechLMScore (r=0.0305, p=0.6741) exhibit the lowest correlation. SpeechLMScore's low influence of noise, along with its high correlation with the severity scores confirms its robustness.

Not surprisingly, WADA SNR shows the highest absolute correlation (r=-0.2461, p < 0.001). Voicing percentage (r=-0.1708, p < 0.05), CPP (r= 0.1596 p < 0.05), Shimmer (r=0.1620, p < 0.05), and Phoneme Error Rate (r = 0.1459, p < 0.05) also show weak correlations. It is well known that pitch estimation is very sensitive to noise, which explains the pitch-based parameters reliance. As CPP is known to be a robust feature, it's sensitiveness is somewhat surprising.

\subsection{Limitations and plan for future work}

In this work, we only made a preliminary investigation of the SpeechLMScore for this task. We expect that retraining the LSTM language model with more relevant healthy data for the task will improve performance, such as large Dutch typical speech. 
However, such a language model requires approximately 50k hours of data. We would like to also broaden our analysis to other Dutch datasets such as the COPAS \cite{middag2012automatic}, and the NKI-RUG-UMCG \cite{halpern2022manipulation}. Other improvements can come from the investigation of different self-supervised features such as wav2vec \cite{schneider2019wav2vec}or WavLM \cite{chen2022wavlm}, and the selection of appropriate layers. For now, it is still expected that language-specific phonetic information will outperform self-supervised acoustic units so, we also plan to use phonetic posteriorgram features in future comparison.

An obvious disadvantage of SpeechLMScore in comparison to the acoustic feature approaches is the lack of interpretability. We think that this can be possibly overcome by interpreting the acoustic units discovered by self-supervised features.

\section{Conclusion}

This study investigated the effectiveness of a reference-free speech severity evaluation method, SpeechLMScore, in comparison to traditional acoustic feature-based approaches and a reference-based phoneme error rate (PER) model. Our results across both the NKI-SpeechRT and NKI-OC-VC datasets consistently demonstrated the superior performance of SpeechLMScore over individual acoustic features, which have historically been used for speech pathology evaluation. Our findings also show SpeechLMScore’s robustness to noise, a common issue in real-world speech datasets, particularly in clinical recordings. Future work should focus on exploring different self-supervised feature performance settings and improving the interpretability of the model which is important for clinical use.

\begin{acks}

The authors would like to thank Thomas Tienkamp for his extensive comments on the analysis. The data collection in the paper received ethical approval under the numbers IRBd20-159 (NKI-OC-VC), IRBd19-025 and N05TSP (NKI-SpeechRT). This work is partly financed by the Dutch Research Council (NWO) under project number 019.232SG.011 titled "I don't sound like myself": Creating voice conversion-based speech
technology for healthcare", and partly supported by JST CREST JPMJCR19A3, Japan.

\end{acks}
\bibliographystyle{ACM-Reference-Format}
\bibliography{sample-sigconf}

\end{document}